# Trust-Aware Sybil Attack Detection for Resilient Vehicular Communication

Thomas L. Morton[1], Abinash Borah[1], Anirudh Paranjothi[1]

[1]Department of Computer Science, Oklahoma State University, Stillwater Oklahoma, USA

**Correspondence**
Thomas L. Morton, Department of Computer Science, Oklahoma State University, Stillwater, Oklahoma, USA
Email: thomas.morton@okstate.edu
ORCID ID: 0009-0000-0774-6524

**ABSTRACT**
Connected autonomous vehicles, or Vehicular Ad hoc Networks (VANETs), hold great promise, but concerns persist regarding safety, privacy, and security, particularly in the face of Sybil attacks, where malicious entities falsify neighboring traffic information. Despite advancements in detection techniques, many approaches suffer from processing delays and reliance on broad architecture, posing significant risks in mitigating attack damages. To address these concerns, our research proposes a Trust Aware Sybil Event Recognition (TASER) framework for assessing the integrity of vehicle data in VANETs. This framework evaluates information exchanged within local vehicle clusters, maintaining a cumulative trust metric for each vehicle based on reported data integrity. Suspicious entities failing to meet trust metric thresholds are statistically evaluated, and their legitimacy is challenged using directional antennas to verify their reported GPS locations. We evaluate our framework using the OMNeT++ discrete event simulator, SUMO traffic simulator, and VEINS interface with TraCI API. Our approach reduces attack detection times by up to 66% in urban scenarios, with accuracy varying by no more than 3% across simulations containing up to 30% Sybil nodes and operates without reliance on roadside infrastructure.

Keywords: Autonomous Vehicles, VANET, Sybil Attacks, Trust Metrics, Wireless Networks, Wireless Security, Directional Antennas

## 1. INTRODUCTION

Autonomous vehicular technology stands at the forefront of innovation, offering significant time and cost savings for individuals and industries. However, its success depends on an effective framework for collaboration among vehicles and their surrounding infrastructure. To meet this requirement, the concept of vehicular ad hoc networks (VANETs) has been developed, utilizing networking principles derived from current wireless local area network (WLAN) technology.

VANET architecture consists of three communication domains: vehicle-to-vehicle (V2V) communications, vehicle-to-infrastructure (V2I) communications, and infrastructure-to-infrastructure (I2I) communications, shown in Figure 1. These domains, supported by specialized hardware, establish a highly responsive network that delivers real-time updates on traffic conditions, hazards, and emergency vehicle movements, along with supplementary services such as weather data, pedestrian traffic monitoring, and advertising for local attractions. As such, VANETs drive advancements in autonomous technology by ensuring safe commuting conditions and providing novel services to passengers.

VANETs inherit vulnerabilities from WLAN networks while also encountering unique challenges due to their dynamic nature[1,2]. A primary threat is the Sybil attack, where malicious actors create multiple fake identities within the network, distorting legitimate traffic information and opening pathways for advanced attacks such as denial-of-service, blackhole, and man-in-the-middle attacks[3,4]. These attacks compromise the network's integrity and undermine functionality, disrupt trust, and weaken consensus mechanisms. By generating Sybil vehicles, attackers can alter the trajectories of other vehicles[5] and disrupt cooperative driving algorithms, potentially leading to road collisions[6].

Detecting Sybil attacks in VANETs presents significant challenges due to frequent network topology changes and the sensitivity of commuter traffic data[7,8]. Effective detection mechanisms must be both responsive and privacy-preserving, with the capacity to function independently of critical infrastructure. To meet these demands, we propose a Trust Aware Sybil Event Recognition (TASER) framework, whereby each vehicle maintains a table of reputation scores for neighboring vehicles based on observed behaviors. TASER effectively mitigates Sybil threats even in regions lacking Internet access or where installation and maintenance of roadside infrastructure is prohibitively costly.

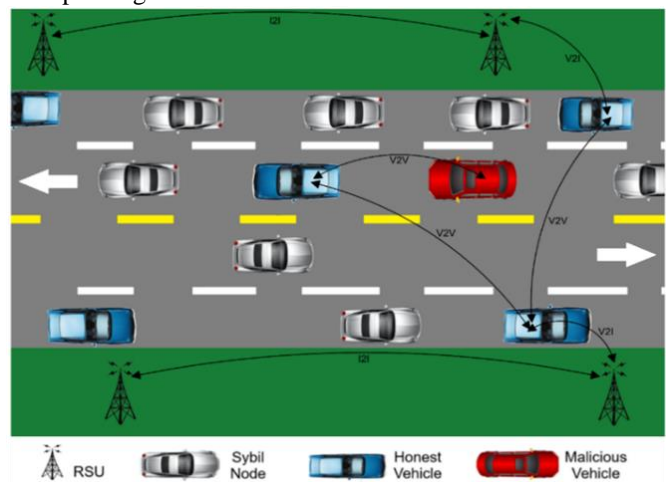

FIGURE 1 - Sybil Attack and Communication Domains in VANET

Our framework contributes to the literature by enabling the classification of potentially malicious nodes within VANETs without relying on external Internet connectivity or infrastructure beyond V2V communication. The TASER framework introduces a lightweight, rapid-response mechanism for Sybil event detection that operates autonomously, eliminating the need for intervention from governing agencies or certifying authorities. The novelty of TASER lies in its dual approach: the use of directional antennas combined with the absence of roadside infrastructure. Directional antennas improve detection accuracy by reducing false positives and alleviating network congestion through localized validation requests, while the lack of roadside infrastructure facilitates cost-effective deployment in remote areas where traditional infrastructure may be financially or logistically unfeasible.

To provide a comprehensive understanding of our research, we structure this paper as follows: Section II reviews related work on Sybil attack detection in VANETs. Section III describes our framework's trust metrics for rogue node detection. Section IV evaluates the effectiveness of our framework through simulated traffic scenarios. Finally, Section V presents our conclusions and discusses future work.

## 2. LITERATURE ON SYBIL ATTACK DETECTION

This section presents an overview of some recent approaches to Sybil attack detection which loosely fall into three categories: encryption & authentication schemes, identity management schemes, and message anomaly analysis[9].

Encryption and authentication implementations ensure the integrity and non-repudiation of entities and messages. To exemplify, Funderburg and Lee[10] propose a trusted authority (TA) registration scheme, where vehicles obtain group keys for message signing and pseudonym generation. Vehicles register their true identities with the TA, which then assigns them to a group, requiring all communications to be signed with group signatures to enhance sender authentication among neighboring vehicles. Baza et al.[11] utilize a Proof-of-Work algorithm and encryption for trajectory authentication, where vehicles solve a puzzle upon encountering a roadside unit (RSU) to validate their route. Mulambia, Varshney, and Suman[12] introduce a blockchain-based approach to vehicle validation in which vehicles authenticate with multiple RSUs within a decentralized peer-to-peer network, facilitating secure authentication and verification of vehicle locations. While these methods provide strong defenses against Sybil attacks, they can be computationally intensive and often depend on a central authority for credential distribution and verification.

Identity management implementations associate a concrete identity with each vehicle, allowing its activities to be attributed to a responsible party and enabling the revocation of communication capabilities in cases of malicious behavior. For example, Hamdan, Hudaib, and Awajan[13] implement identity management through pseudonyms issued by a TA, such as the Department of Motor Vehicles (DMV), which generate coarse-grained hashes for vehicle identification while preserving anonymity. In cases of suspicious activity, hashes are sent to the TA to verify if the identities belong to the same fine-grained group, indicating an attack. Similarly, Qi et al.[2] adopt an identity-based approach through a TA responsible for maintaining vehicle trust data and managing certificates. Hasrouny et al.[14] utilize local groups where vehicles track the identities of their neighbors, requiring registration with the DMV for a signed certificate from a certificate authority (CA). Vehicles share trust observations within their broadcast range, which are periodically sent to RSUs for analysis and malicious classification. While these implementations detect attacks by limiting the pool of forgeable identities, they often require a central credential distribution and verification authority. Additionally, requiring registration and validation with unique identities risks commuter privacy by enabling the tracking of frequently visited locations.

Message anomaly analysis investigates the communications between vehicles and the VANET's supporting infrastructure, as deviations from expected behavior may signify an attack and assist in identifying its type and perpetrator. For instance, Benkirane[15] analyzes data within vehicle broadcast messages to validate legitimacy through GPS location, triangulated using multiple RSUs. Discrepancies between reported locations and RSU observations are deemed suspicious. Similarly, Azam et al.[16] employ a robust anomaly detection mechanism using machine learning classifiers on velocity and other data, implementing hard- and soft-voting systems for malicious activity detection. Rathee et al.[7] apply Dempster-Shafer theory and a recommender algorithm to manage trust metrics by identifying anomalies in energy consumption, communication delays, resource utilization, and multi-hop transmission scenarios. Although anomaly analysis effectively detects irregularities indicative of Sybil attacks, it often requires extensive data collection and significant computational resources.

Our implementation addresses the limitations of these methods by maintaining a trust metric associated with temporary pseudonyms for vehicles. Each vehicle independently creates an identity for each neighboring vehicle. From there, the vehicle computes and maintains a trust metric for all other vehicles within its communication range, eliminating reliance on RSUs or centralized authorities. The TASER framework's lightweight trust computation, based on analysis of the messages exchanged between vehicles, effectively meets the low-latency requirements of VANETs.

# 3. PROPOSED WORK

In this section, we outline the TASER framework that serves as a trust assignment and analysis system for Sybil detection. Our objectives are as follows: to maintain observational data of neighboring vehicles within the network, detect and classify Sybil nodes using this data, achieve accurate classification of Sybil nodes exclusively within the V2V domain, and ensure rapid detection times. We achieve these goals through a combination of trust metrics established from the observation of data from broadcast messages and the validation of suspect vehicles' physical presence through directional messaging. The overview of our detection framework is presented in the flowchart in Figure 2.

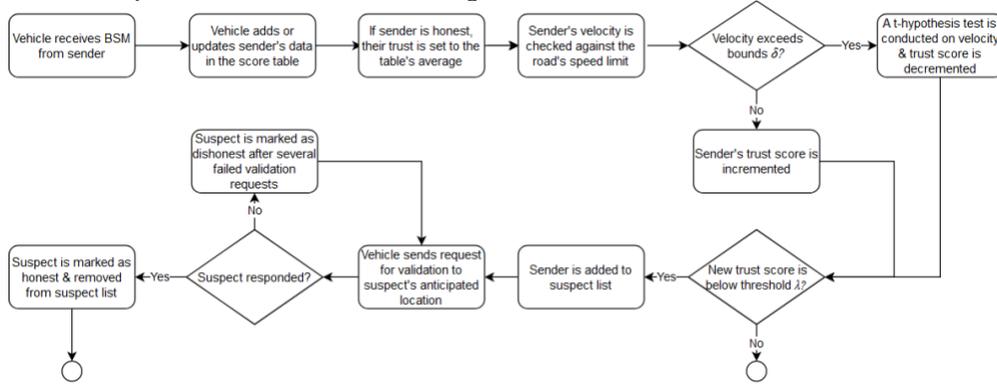

*FIGURE 2 - Flowchart of TASER trust calculation and verification*

## 3.1 Initialization

While in transit, each vehicle creates a table of records, $STable_{self}$, that stores the most recently transmitted status information from the vehicle's neighbors. This table contains the message sender's pseudonym, GPS coordinate data, last reported traveling speed, last recorded message timestamp, classification category, and a trust score used to classify the vehicle as legitimate or malicious. Each vehicle maintains a running average of all trust values within its score table as *AverageTrust*, which is used to initialize new table entries and evaluate neighboring vehicle behaviors.

Each vehicle periodically broadcasts a status message, $BSM_{status}$, to all other vehicles within its transmission range containing the sender's pseudonym, velocity, timestamp, and GPS data. Upon receiving a $BSM_{status}$ message, the recipient references $STable_{self}$ to determine whether the recipient has previously received a $BSM_{status}$ message from the sender. If the recipient's score table does not contain an entry for the message sender, the recipient creates a new entry within $STable_{self}$ to store the data contained within the $BSM_{status}$ message and assigns a score equal to the average of all scores in the table.

After updating the score table with the incoming message data, the reported velocity is compared to the current route's speed limit, denoted as *SpeedLimit*. If the velocity is within a defined threshold of $\pm\delta\%$ of this limit, the sender's trust score in the recipient's score table is incremented. Vehicle velocities typically remain within this range due to physical and regulatory constraints, with drivers making incremental adjustments to match the speeds of neighboring vehicles and maintain safe following distances. Additionally, penalties for both speeding and driving excessively slowly promote compliance with traffic regulations, further minimizing extreme velocity deviations. Observations that exceed the threshold undergo further analysis, detailed in Section 3.2, and trust values for the vehicle are subsequently updated.

In adjusting a vehicle's trust, the observer incorporates historical data to determine the adjustment magnitude, calculated as the previous trust score multiplied by $\beta$. Vehicles with velocities within the threshold receive a standard increment of $1 +$ current trust $\times \beta$, while deductions are computed as $1 -$ current trust $\times \beta$. Consequently, vehicles with low or negative trust values incur larger deductions and smaller increments, whereas those with positive trust values experience the opposite effect. Entities that consistently exhibit suspicious behavior gain trust at a diminished rate, thereby impeding attackers from intermittently providing accurate information to recover credibility. To further limit an attacker's ability to inflate trust values, the minimum and maximum trust scores are bounded at $\pm 5$, with plans for normalization to $\pm 1$ in future research.

## 3.2 Challenging Vehicle Authenticity

When the difference in reported velocity exceeds the dynamic threshold, the vehicle's data is flagged as suspect, prompting further analysis. The observing vehicle then analyzes the received data using a two-tailed t-hypothesis test, wherein the null hypothesis is that the vehicle is legitimate and the alternative hypothesis being that the vehicle is potentially malicious.

For the hypothesis test, the vehicle uses a significance level $\alpha$ with a tight bound, which prioritizes the mitigation of type 2 errors over type 1 errors. Type 1 errors, characterized by falsely classifying honest vehicles as malicious, may lead to occasional unnecessary interventions, such as broadcasting messages to confirm vehicle presence. However, the consequences of type 2 errors are far more severe, resulting in a failure to identify truly malicious vehicles and thus leaving the network vulnerable to undetected attacks. Upon rejecting the null hypothesis, the recipient decrements the sender's trust score within *ScoreTable$_{self}$* using the weighting $\beta$. Each failed evaluation corresponds to an increased likelihood that the sender is malicious.

After updating the trust value associated with the message sender, the recipient then recalculates *AverageTrust* and compares this value to the sender's current trust. If the sender's trust value deviates excessively from the *AverageTrust*, bounded by $\lambda$, the vehicle is classified as malicious, added to a suspect list, and subjected to investigation by directional messaging. The trust assessment method of our framework is presented in Algorithm 1.

---

**Algorithm 1** Trust Assessment for Neighboring Vehicles
---
**Procedure** UpdateScoreEntries(STable, BSM, SpeedLimit)
  AverageTrust ← STable.TableAverageTrust()
  **if** BSM$_{ID}$ ∉ Stable **then**
    STable.Append(BSM$_{ID}$, BSM$_{Speed}$, AverageTrust)
  **if** STable$_{ID}$[TYPE] = "honest" **and** STable$_{ID}$[TRUST] < AverageTrust **then**
    STable$_{ID}$[TRUST] ← AverageTrust
  **if** BSM$_{Speed}$ > SpeedLimit × (1 + $\delta$) **or** BSM$_{Speed}$ < SpeedLimit × (1 – $\delta$) **then**
    **if** THypothesisTest() = *True* **then**
      STable$_{ID}$[TRUST] ← STable$_{ID}$[TRUST] – (1 – $\beta$ × STable$_{ID}$[TRUST])
    **else**
      Vehicle is suspect, but trusted for this iteration – no score change
  **else**
    STable$_{ID}$[TRUST] ← STable$_{ID}$[TRUST] + (1 + $\beta$ × STable$_{ID}$[TRUST])
  **if** AverageTrust – STable$_{ID}$[TRUST] ≥ $\lambda$ × AverageTrust **then**
    AddSuspectList(BSM$_{ID}$)
**End procedure**

*ALGORITHM 1 - Trust assessment algorithm to classify Sybil nodes within a local VANET.*

## 3.3 Confirming Vehicle Legitimacy

When a vehicle is classified as potentially malicious, the classifying vehicle initiates a challenge process by generating a series of challenge packets. These packets contain a pseudo-random challenge number and are addressed to the suspected vehicle. By leveraging coordinate data from the suspect's previous messages, the classifying vehicle calculates the anticipated location of the suspect vehicle and sends a directed message to that location, utilizing existing velocity and trajectory data.

Upon receiving a challenge packet, a legitimate vehicle responds with a corresponding packet containing the same pseudo-random number. This response authenticates the vehicle's physical presence, and the vehicle is marked as honest in the table. However, Sybil nodes, which lack a physical presence, will not be located at the anticipated target location and thus are unable to generate a valid response. The use of directed messaging also mitigates the potential for the attacking vehicle to receive and respond to the challenge. Furthermore, if an attacker attempts to deceive the system by advertising multiple Sybil nodes as occupying the same location, these vehicles will be promptly identified as malicious and classified accordingly.

## 4. PERFORMANCE EVALUATION

In this section, we assess the performance of our framework across various urban scenarios within our simulation environment. We additionally conduct a comparative analysis between our framework and two other established approaches to attack detection: Hybrid Trust Model & Misbehavior Detection System (HTM/MDS)[14] and Host-Based Intrusion Detection System (Host-Based IDS)[17].

### 4.1 Simulation Setup

We assess our detection framework's effectiveness using OMNeT++ 5.7 IDE, Veins 5.2, and SUMO 1.12.0 for simulation and scenario management in a setup featuring two-lane roadways totaling 2 kilometers, with traffic control architecture. Simulations adhere to 802.11p transmission protocols, broadcasting every 100 milliseconds. Each scenario contains 500 vehicles, with 5% to 30% designated as Sybil nodes, generated via seeded random number allocation. Honest nodes' velocities are set at 15 meters per second, while Sybil nodes travel at 2 meters per second. Algorithm parameters are set as follows: $\alpha$ = 0.01, $\beta$ = 0.1, $\delta$ = 0.4, and $\lambda$ = 0.15 and exhibit the following behavior: $\alpha$ affects the sensitivity to which our t-hypothesis test rejects the null hypothesis. $\beta$ acts as a weighting parameter for trust evaluations; larger values for $\beta$ amplify the influence of prior trust observations on updating trust values. Greater $\delta$ values expand the range of permissible velocities. $\lambda$ gives the threshold for inadequate trust values; smaller values for $\lambda$ reduce the permissible range for vehicles exhibiting low trust.

### 4.2 Performance Metrics

Our framework's performance is evaluated using detection times along with accuracy, F1-Score, and specificity to demonstrate variance in our framework's false positive rate (FPR) for different $\delta$ values in our algorithm. Accuracy represents the percentage of correctly identified nodes, specificity details the degree to which our algorithm does not encounter false positives, and F1 Score describes a harmonic balance between accurate classification of true positives and true negatives among their respective categories[18]. Additionally, we define classification times as the period between the initial encounter and final classification, measured in epochs.

## 4.3 Framework Efficacy

**Accuracy:** In Figure 3(a), we observe a high degree of accuracy across all simulation scenarios, with only a marginal loss as the number of malicious nodes increases. Furthermore, the TASER framework exhibits stability in its accuracy, demonstrating a variance of approximately 3%. This stands in contrast to our comparative frameworks, which present variances of 17% and 30% for HTM/MDS and Host-Based IDS, respectively. This observation highlights the robustness of the TASER framework, as it maintains accurate classifications irrespective of the initial density of encountered malicious nodes.

**F1 Score:** In Figure 3(b), our framework presents F1 scores that match or exceed those of our comparative frameworks, indicating an ability to accurately classify both true positives and true negatives during evaluation. While the F1 score evaluation of the TASER framework shows some instability across varying densities of malicious nodes, we contend that this behavior is due to the dynamic nature of traffic patterns relative to the velocity data used during evaluation. Despite this variability, our framework consistently achieves an F1 score that surpasses comparative frameworks across all scenarios by an approximate average of 7% for Host-Based IDS and 15% for HTM/MDS.

**Detection Times:** Our framework demonstrates exceptional performance in detection times, achieving an approximate two- to three-fold reduction over our comparative frameworks, as depicted in Figure 3(c). Furthermore, the TASER framework maintains stable detection times even under increased densities of Sybil nodes. This observation is particularly significant given the incremental growth in detection times observed in our comparative frameworks, with HTM/MDS exhibiting more than a threefold increase in time to final classification under greater densities of malicious nodes.

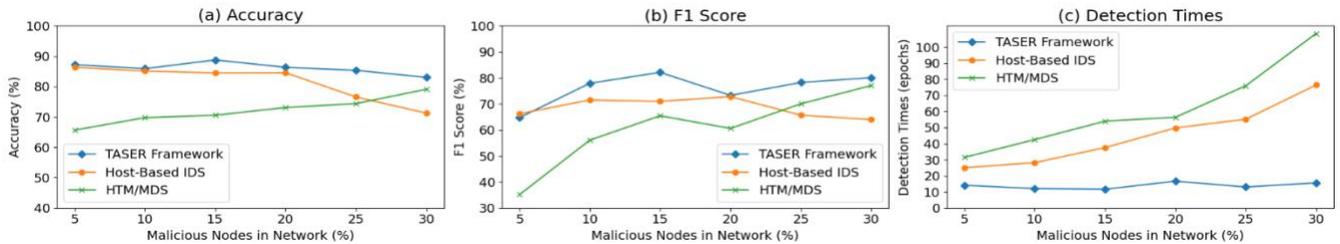

*FIGURE 3 - Comparison of Accuracy, F1 Score, and Detection Metrics for the TASER framework*

Table 1 further details the TASER framework's performance, demonstrating consistent accuracy, efficient detection times, and robust F1 scores, with stable performance across increasing malicious node densities. This stability is enhanced by the framework's adaptive $\beta$ trust-weighting and the minimum trust threshold modifier $\lambda$. In low attack densities, Sybil nodes incur heavier penalties for repeated suspicious reporting, while honest nodes maintain high average trust across all densities, preserving trust for legitimate participants. By dynamically adjusting detection thresholds according to network trust levels, the framework prompts earlier validation requests as communal trust decreases.

| TASER Performance Metrics | | | |
|---|---|---|---|
| | Accuracy | F1 Score | Time to Detection |
| **5% Sybil Nodes** | 87.16% | 64.72% | 14.14 epochs |
| **10% Sybil Nodes** | 88.77% | 77.89% | 12.08 epochs |
| **15% Sybil Nodes** | 86.31% | 82.08% | 11.68 epochs |
| **20% Sybil Nodes** | 88.72% | 73.25% | 16.68 epochs |
| **25% Sybil Nodes** | 85.32% | 78.2% | 13.08 epochs |
| **30% Sybil Nodes** | 82.99% | 80.05% | 15.6 epochs |

*TABLE 1 - TASER performance metrics*

During testing, we further investigate the impact of various $\lambda$ values, which directly influence the threshold at which our algorithm identifies vehicles as potentially malicious. We observe that smaller $\lambda$ values correlate with a reduction in detection times; however, this trend was accompanied by an increase in false positive rates, as illustrated in Figures 4(a) and 4(b).

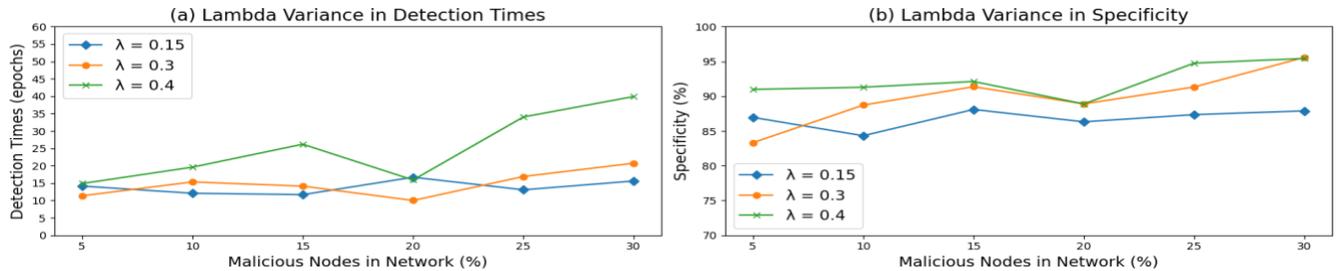

*FIGURE 4 - Analysis of influence on detection times and specificity for values of lambda*

We attribute this observation to larger values of λ increasing the minimum detection threshold, thus requiring a larger difference between individual vehicles' trust and the *AverageTrust* for a given vehicle's table. We contend that the improvements to specificity presented by larger λ values do not justify the increase in detection times, given our framework's relative ease in adapting to false positives. As such, we consider λ values between 0.15 and 0.3 to be sufficient.

## 5. CONCLUSIONS AND FUTURE WORK

In this paper, we present a robust approach to Sybil attack detection by utilizing a trust-scoring metric based on vehicles' observations of neighboring vehicles' traffic patterns. We consider detection solely through the V2V domain and without mediation through VANET infrastructure or governing bodies. Vehicles that provoke suspicion within the network are sent a challenge packet via a directional antenna which is used to authenticate their physical presence. Our proposed framework drastically reduces detection times for Sybil attacks in VANETs in urban scenarios while maintaining a high degree of accuracy and minimal false positives in detection. The contributions of our framework stand to reduce the risks of potentially fatal outcomes from Sybil attacks in VANETs, regardless of whether broad network architecture is available.

Future implementations will integrate additional data collected at the vehicular level to enhance our algorithm's robustness and stability. We also intend to investigate sharing trust observations between vehicles, potentially reducing response times in attack detection. Furthermore, we propose integration with existing identity-management systems to establish an effective reporting mechanism for malicious actors when a vehicle establishes connections with the broader VANET infrastructure.

## FINANCIAL DISCLOSURE
This research was funded by the Advancing Undergraduate Research and Creative Activity (AURCA) program at Oklahoma State University. We gratefully acknowledge their contributions to our research in autonomous vehicle technology.

## REFERENCES
1. Syla V, Lala A, Biberaj A. Vanet Security and Privacy—An Overview. In: EIRP Proceedings. 2024; 19.
2. Qi J, Zheng N, Xu M, Chen P, Li W. A Hybrid-Trust-Based Emergency Message Dissemination Model for Vehicular Ad Hoc Networks. Journal of Information Security and Applications. 2024;81.
3. Arif M, Wang G, Bhuiyan MZA, Wang T, Chen J. A Survey on Security Attacks in VANETs: Communication, Applications and Challenges. Vehicular Communications. 2019;19.
4. Rakhi S, Shobha KR. LCSS based SYBIL Attack Detection and Avoidance in Clustered Vehicular Networks. IEEE Access. 2023;11.
5. Zhang Z, Yingxu L, Ye C, Jingwen W, Yuhang W. Detection Method to Eliminate Sybil Attacks in Vehicular Ad-hoc Networks. Ad Hoc Networks. 2023; 141.
6. Tulay HB, Koksal CE. Sybil Attack Detection based on Signal Clustering in Vehicular Networks. IEEE Transactions on Machine Learning in Communications and Networking. 2024; 2.
7. Rathee G, Kumar A, Kerrache CA, Calafate C. A Trust Management Solution for 5G-based Future Generation Internet of Vehicles. Computer Networks. 2024;248.
8. Sohail et al. Routing Protocols in Vehicular Adhoc Networks (VANETs): A Comprehensive Survey. Internet of Things. 2023;23.
9. Park S, Aslam B, Turgut D, Zou C. Defense Against Sybil Attack in Vehicular Ad Hoc Network Based on Roadside Unit Support. In: Military Communications Conference. 2009.
10. Funderburg LE, Lee I-Y. A Privacy-Preserving Key Management Scheme with Support for Sybil Attack Detection in VANETs. Sensors. 2021;21.
11. Baza M, Nabil M, Mahmoud MMEA, et al. Detecting Sybil Attacks Using Proofs of Work and Location in VANETs. IEEE Transactions on Dependable and Secure Computing. 2022;19.
12. Mulambia C, Varshney S, Suman A. Privacy Preserving Blockchain Based Authentication Scheme for VANET. Engineered Science. 2024;28.
13. Hamdan S, Hudaib A, Awajan A. Hybrid Algorithm to Detect the Sybil Attacks in VANET. In: 2018 Fifth International Symposium on Innovation in Information and Communication Technology (ISIICT). 2018.
14. Hasrouny H, Samhat AE, Bassil C, Laouiti A. Trust Model for Secure Group Leader-Based Communications in VANET. Wireless Networks. 2019;25.
15. Benkirane S. Road Safety Against Sybil Attacks Based on RSU Collaboration in VANET Environment. In: 5th International Conference, MSPN, Mohammedia, Morocco. 2019.
16. Azam S, Bibi M, Riaz R, Rizvi SS, Kwon SJ. Collaborative Learning Based Sybil Attack Detection in Vehicular Ad-Hoc Networks (VANETS). Sensors. 2022;22.
17. Zaidi K, Milojevic MB, Rakocevic V, Nallanathan A, Rajarajan M. Host-Based Intrusion Detection for VANETs: A Statistical Approach to Rogue Node Detection. IEEE Transactions on Vehicular Technology. 2016;65.
18. Hossin M, Sulaiman MN. A Review on Evaluation Metrics for Data Classification Evaluations. International Journal of Data Mining & Knowledge Management Process (IJDKP). 2015;5.